\documentclass[a4paper,10pt, twocolumn]{article}

\usepackage[margin=2.0cm]{geometry}

\usepackage{calc}
\usepackage{enumitem}

\usepackage[american]{babel}
\usepackage{csquotes}

\usepackage{amsmath}

\usepackage[utf8]{inputenc}

\usepackage{tikz,lipsum,lmodern}
\usepackage[most]{tcolorbox}

\usepackage[ttscale=.875]{libertine}
\usepackage[scaled=0.96]{zi4}

\usepackage{nicefrac}

\usepackage[backend=biber,
            style=apa,
            apamaxprtauth=99,
            natbib=true]{biblatex}
\DeclareLanguageMapping{american}{american-apa}

\usepackage{graphicx}

\usepackage[labelsep=period]{caption}

\usepackage{xcolor}
\definecolor{blendedblue}{rgb}{0.2, 0.2, 0.6}
\definecolor{blendedred}{rgb}{0.8, 0.2, 0.2}
\usepackage[bookmarks=true,
            breaklinks=true,
            pdfborder={0 0 0},
            citecolor=blendedblue,
            colorlinks=true,
            linkcolor=blendedblue,
            urlcolor=blendedblue,
            citecolor=blendedblue,
            linktocpage=false,
            hyperindex=true,
            linkbordercolor=white]{hyperref}
\hypersetup{colorlinks=true}

\bibliography{draft.bib}

\usepackage{fancyhdr}
\makeatletter
\renewcommand{\maketitle}{\bgroup\setlength{\parindent}{0pt}
\begin{flushleft}
  \textbf{\huge\@title\\}
  \vspace{5mm}
  \@author
\end{flushleft}\egroup
}
\makeatother

\usepackage{tikz}
\usetikzlibrary{positioning,shapes,shadows,arrows}

\title{Sustainable computational science:\\
       the ReScience initiative}
\author{%
  \begin{small}
    \textbf{Nicolas P. Rougier}$^{1\dagger\S\star}$\href{http://orcid.org/0000-0002-6972-589X}{\includegraphics[width=8pt]{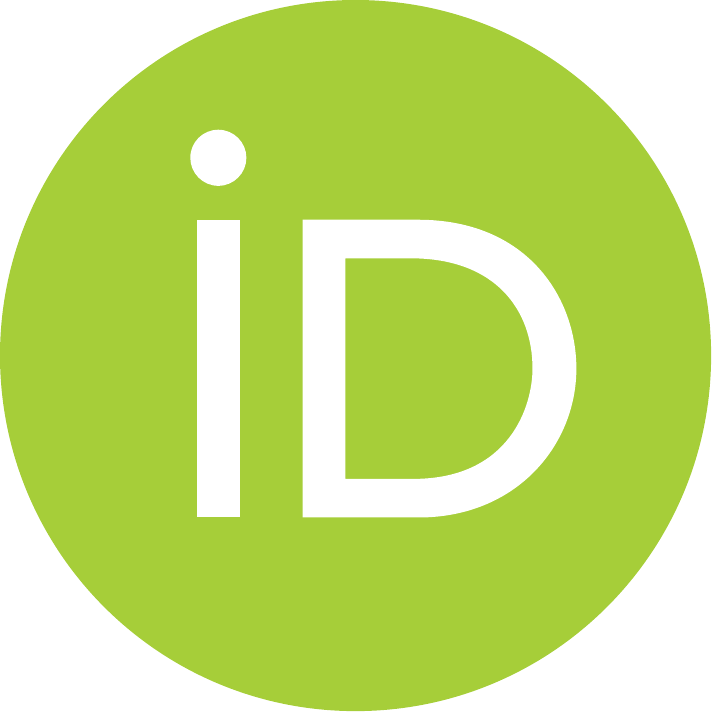}},
\textbf{Konrad Hinsen}$^{2\dagger\star}$\href{http://orcid.org/0000-0003-0330-9428}{\includegraphics[width=8pt]{orcid}},
\textbf{Frédéric Alexandre}$^{1\S}$\href{http://orcid.org/0000-0002-6113-1878}{\includegraphics[width=8pt]{orcid}},
\textbf{Thomas Arildsen}$^{3\dagger}$\href{http://orcid.org/0000-0003-3254-3790}{\includegraphics[width=8pt]{orcid}},
\textbf{Lorena Barba}$^{4\dagger}$\href{http://orcid.org/0000-0001-5812-2711}{\includegraphics[width=8pt]{orcid}},
\textbf{Fabien C. Y. Benureau}$^{1\ddagger}$\href{http://orcid.org/0000-0003-4083-4512}{\includegraphics[width=8pt]{orcid}},
\textbf{C. Titus Brown}$^{5\dagger}$\href{http://orcid.org/0000-0001-6001-2677}{\includegraphics[width=8pt]{orcid}},
\textbf{Pierre de Buyl}$^{6\dagger}$\href{http://orcid.org/0000-0002-6640-6463}{\includegraphics[width=8pt]{orcid}},
\textbf{Ozan Caglayan}$^{7\ddagger}$\href{http://orcid.org/0000-0002-5992-3470}{\includegraphics[width=8pt]{orcid}},
\textbf{Andrew P. Davison}$^{8\ddagger}$\href{http://orcid.org/0000-0002-4793-7541}{\includegraphics[width=8pt]{orcid}},
\textbf{Marc André Delsuc}$^{9\ddagger}$\href{http://orcid.org/0000-0002-1400-5326}{\includegraphics[width=8pt]{orcid}},
\textbf{Georgios Detorakis}$^{10\ddagger\S}$\href{http://orcid.org/0000-0001-5891-1702}{\includegraphics[width=8pt]{orcid}},
\textbf{Alexandra K. Diem}$^{11\S}$\href{http://orcid.org/0000-0003-1719-1942}{\includegraphics[width=8pt]{orcid}},
\textbf{Damien Drix}$^{12\ddagger}$\href{http://orcid.org/0000-0003-4107-5693}{\includegraphics[width=8pt]{orcid}},
\textbf{Pierre Enel}$^{13\ddagger}$\href{http://orcid.org/0000-0001-8983-6223}{\includegraphics[width=8pt]{orcid}},
\textbf{Benoît Girard}$^{14\dagger\ddagger\S}$\href{http://orcid.org/0000-0002-8117-7064}{\includegraphics[width=8pt]{orcid}},
\textbf{Olivia Guest}$^{15\dagger}$\href{http://orcid.org/0000-0002-1891-0972}{\includegraphics[width=8pt]{orcid}},
\textbf{Matt G. Hall}$^{16\ddagger}$\href{http://orcid.org/0000-0002-9530-5477}{\includegraphics[width=8pt]{orcid}},
\textbf{Rafael Neto Henriques}$^{17\S}$,
\textbf{Xavier Hinaut}$^{1\ddagger}$\href{http://orcid.org/0000-0002-1924-1184}{\includegraphics[width=8pt]{orcid}},
\textbf{Kamil S Jaron}$^{18\ddagger}$\href{http://orcid.org/0000-0003-1470-5450}{\includegraphics[width=8pt]{orcid}},
\textbf{Mehdi Khamassi}$^{14\ddagger\S}$\href{http://orcid.org/0000-0002-2515-1046}{\includegraphics[width=8pt]{orcid}},
\textbf{Almar Klein}$^{19\ddagger}$\href{http://orcid.org/0000-0002-9978-2780}{\includegraphics[width=8pt]{orcid}},
\textbf{Tiina Manninen}$^{20\ddagger}$\href{http://orcid.org/0000-0002-0456-1185}{\includegraphics[width=8pt]{orcid}},
\textbf{Pietro Marchesi}$^{21\ddagger}$\href{http://orcid.org/0000-0001-5955-6909}{\includegraphics[width=8pt]{orcid}},
\textbf{Dan McGlinn}$^{22\ddagger}$\href{http://orcid.org/0000-0003-2359-3526}{\includegraphics[width=8pt]{orcid}},
\textbf{Christoph Metzner}$^{23\ddagger\S}$\href{http://orcid.org/0000-0002-5933-2101}{\includegraphics[width=8pt]{orcid}},
\textbf{Owen L. Petchey}$^{24\S}$\href{http://orcid.org/0000-0002-7724-1633}{\includegraphics[width=8pt]{orcid}},
\textbf{Hans Ekkehard Plesser}$^{25\ddagger}$\href{http://orcid.org/0000-0001-7843-5993}{\includegraphics[width=8pt]{orcid}},
\textbf{Timothée Poisot}$^{26\dagger}$\href{http://orcid.org/0000-0002-0735-5184}{\includegraphics[width=8pt]{orcid}},
\textbf{Karthik Ram}$^{27\dagger}$\href{http://orcid.org/0000-0002-0233-1757}{\includegraphics[width=8pt]{orcid}},
\textbf{Yoav Ram}$^{28\ddagger}$\href{http://orcid.org/0000-0002-9653-4458}{\includegraphics[width=8pt]{orcid}},
\textbf{Etienne Roesch}$^{29\ddagger}$\href{http://orcid.org/0000-0002-8913-4173}{\includegraphics[width=8pt]{orcid}},
\textbf{Cyrille Rossant}$^{30\ddagger}$\href{http://orcid.org/0000-0003-2069-9093}{\includegraphics[width=8pt]{orcid}},
\textbf{Vahid Rostami}$^{31\S}$\href{http://orcid.org/0000-0002-3851-0220}{\includegraphics[width=8pt]{orcid}},
\textbf{Aaron Shifman}$^{32\ddagger\S}$\href{http://orcid.org/0000-0003-2140-7590}{\includegraphics[width=8pt]{orcid}},
\textbf{Joseph Stachelek}$^{33\S}$\href{http://orcid.org/0000-0002-5924-2464}{\includegraphics[width=8pt]{orcid}},
\textbf{Marcel Stimberg}$^{34\ddagger}$\href{http://orcid.org/0000-0002-2648-4790}{\includegraphics[width=8pt]{orcid}},
\textbf{Frank Stollmeier}$^{35\S}$\href{http://orcid.org/0000-0003-4858-0895}{\includegraphics[width=8pt]{orcid}},
\textbf{Federico Vaggi}$^{36\ddagger}$\href{http://orcid.org/0000-0001-8100-158X}{\includegraphics[width=8pt]{orcid}},
\textbf{Guillaume Viejo}$^{37\S}$,
\textbf{Julien Vitay}$^{38\ddagger\S}$\href{http://orcid.org/0000-0001-5229-2349}{\includegraphics[width=8pt]{orcid}},
\textbf{Anya Vostinar}$^{39\ddagger}$\href{http://orcid.org/0000-0001-7216-5283}{\includegraphics[width=8pt]{orcid}},
\textbf{Roman Yurchak}$^{40\ddagger}$\href{http://orcid.org/0000-0002-2565-4444}{\includegraphics[width=8pt]{orcid}},
\textbf{Tiziano Zito}$^{41\dagger}$\\

  \end{small}
\begin{footnotesize}
  \vspace{2mm}
  $^{1}$INRIA Bordeaux Sud-Ouest Talence, France – Institut des Maladies Neurodégénératives, Université de Bordeaux, CNRS UMR 5293, Bordeaux, France – LaBRI, Université de Bordeaux, Bordeaux INP, CNRS UMR 5800, Talence, France
$^{2}$Centre de Biophysique Moléculaire, CNRS UPR4301, Orléans, France -- Synchrotron SOLEIL, Division Expériences, Gif sur Yvette, France
$^{3}$Department of Electronic Systems, Technical Faculty of IT and Design, Aalborg University, Denmark
$^{4}$Department of Mechanical and Aerospace Engineering, the George Washington University, Washington DC, USA
$^{5}$Department of Population Health and Reproduction, University of California Davis, Davis, CA, USA
$^{6}$Instituut voor Theoretische Fysica, KU Leuven, Belgium -- PdB is a postdoctoral fellow of the Research Foundation -- Flanders (FWO)
$^{7}$Laboratoire d'Informatique de l'Université du Maine (LIUM), Le Mans, France
$^{8}$Unité de Neurosciences, Information et Complexité, CNRS FRE 3693, Gif sur Yvette, France
$^{9}$Institut de Génétique et de Biologie Moléculaire et Cellulaire, INSERM U964, CNRS UMR 7104, Université de Strasbourg, Illkirch, France
$^{10}$Department of Cognitive Sciences, University of California Irvine, USA
$^{11}$Computational Engineering and Design, Faculty of Engineering and the Environment, University of Southampton, UK
$^{12}$Department of Computer Science, Humboldt-Universität zu Berlin
$^{13}$Friedman Brain Institute, Icahn School of Medicine at Mount Sinai, NY, USA
$^{14}$Institut des Systèmes Intelligent et de Robotique, Sorbonne Universités, UPMC Univ Paris 06, CNRS, Paris, France
$^{15}$Experimental Psychology, University College London, UK
$^{16}$UCL Great Ormond St Institute of Child Health. University College London, London, UK
$^{17}$MRC Cognition and Brain Sciences Unit, Cambridge, UK
$^{18}$Department of Ecology and Evolution, University of Lausanne, Lausanne, Switzerland
$^{19}$Independent scholar, Enschede, The Netherlands
$^{20}$BioMediTech Institute and Faculty of Biomedical Sciences and Engineering, Tampere University of Technology, Tampere, Finland
$^{21}$Swammerdam Institute for Life Sciences, Center for Neuroscience, Faculty of Science, University of Amsterdam, Amsterdam, the Netherlands
$^{22}$Department of Biology, College of Charleston, Charleston, SC, USA
$^{23}$Centre for Computer Science and Informatics Research, University of Hertfordshire, UK
$^{24}$Department of Evolutionary Biology and Environmental Studies, University of Zurich, Switzerland
$^{25}$Faculty of Science and Technology, Norwegian University of Life Sciences, Aas, Norway -- Institute of Neuroscience and Medicine (INM-6), Jülich Research Centre, Jülich, Germany
$^{26}$Département de Sciences Biologiques, Université de Montréal, Montréal, Canada
$^{27}$Berkeley Institute for Data Science, University of California Berkeley, Berkeley, CA, USA
$^{28}$Department of Biology, Stanford University, Stanford, CA, USA
$^{29}$Centre for Integrative Neuroscience, School of Psychology, University of Reading, UK
$^{30}$Institute of Neurology, University College London, UK
$^{31}$Institute of Neuroscience \& Medicine (INM-6) and Institute for Advanced Simulation (IAS-6) -- JARA-Brain Institute I (INM-10), Jülich Research Center, Jülich, Germany
$^{32}$Department of Biology, University of Ottawa, Ottawa, Ontario, Canada
$^{33}$Department of Fisheries and Wildlife, Michigan State University, MI, USA
$^{34}$Sorbonne Universités, UPMC Univ Paris 06, INSERM, CNRS, Institut de la Vision, Paris, France
$^{35}$Network Dynamics, Max Planck Institute for Dynamics and Self-Organization, Germany
$^{36}$Amazon, Seattle, USA
$^{37}$Montreal Neurological Institute, McGill University, Montreal, Canada
$^{38}$Professorship for Artificial Intelligence, Department of Computer Science, Chemnitz University of Technology, Chemnitz, Germany
$^{39}$Department of Computer Science, Grinnell College, IA, USA
$^{40}$Symerio, Palaiseau, France
$^{41}$Neural Information Processing Group, University of Tübingen, Germany
\par
  \vspace{2mm}
  $^{\star}$ReScience co-founder
  $^{\dagger}$ReScience editor
  $^{\ddagger}$ReScience reviewer
  $^{\S}$ReScience author\\
  Corresponding author:
            \href{mailto:Nicolas.Rougier@inria.fr}{Nicolas.Rougier@inria.fr}          
  \end{footnotesize}
}

\date{}

\begin{document}

\twocolumn[
  \maketitle
  \vspace{5mm}
  \noindent
  \textbf{Abstract.}
  Computer science offers a large set of tools for prototyping, writing,
  running, testing, validating, sharing and reproducing results, however
  computational science lags behind.  In the best case, authors may provide
  their source code as a compressed archive and they may feel confident their
  research is reproducible. But this is not exactly true. Jonathan Buckheit and
  David Donoho proposed more than two decades ago that an article about
  computational results is advertising, not scholarship. The actual scholarship
  is the full software environment, code, and data that produced the
  result. This implies new workflows, in particular in peer-reviews. Existing
  journals have been slow to adapt: source codes are rarely requested, hardly
  ever actually executed to check that they produce the results advertised in
  the article.
  ReScience is a peer-reviewed journal that targets computational research and
  encourages the explicit replication of already published research, promoting
  new and open-source implementations in order to ensure that the original
  research can be replicated from its description. To achieve this goal, the
  whole publishing chain is radically different from other traditional
  scientific journals. ReScience resides on GitHub where each new
  implementation of a computational study is made available together with
  comments, explanations, and software tests.\\

  \textbf{Keywords:} Open Science, Computational Science,
                     Reproducibility, Replicability, Journal, Publication
  \vfill
]

\clearpage
\section*{Introduction}
There is a replication crisis in Science \citep{Baker:2016, Munaf:2017}. This
crisis has been highlighted in fields as diverse as medicine
\citep{ioannidis:2005}, psychology \citep{nosek:2015}, the political sciences
\citep{janz:2015}, and recently in the biomedical sciences
\citep{iqbal:2016}. The reasons behind such non-replicability are as diverse as
the domains in which it occurs.  In medicine, factors such as {\em study power
  and bias, the number of other studies on the same question, and importantly,
  the ratio of true to no relationships among the all relationships probed}
have been highlighted as important causes \citep{ioannidis:2005}.  In
psychology, non-replicability has been blamed on spurious p-values (p-hacking),
while in the biomedical sciences \citep{iqbal:2016}, a {\em lack of access to
  full datasets and detailed protocols for both clinical and non-clinical
  biomedical investigation} is seen as a critical factor.  The same remarks
were recently issued for chemistry \citep{coudert:2017}.  Surprisingly, the
computational sciences (in the broad sense) and computer sciences (in the
strict sense) are no exception \citep{donoho:2009,Manninen:2017} despite the
fact they rely on code and data rather than on experimental observations, which
should make them immune to the aforementioned problems.\\

When Collberg and colleagues \citep{collberg:2014, collberg:2015} decided to measure the extent of the problem
precisely, they investigated the availability of code and data as well as {\em the extent to which this code would actually build with reasonable
  effort}. The results were dramatic: of the 515 (out of 613) potentially reproducible
papers targeted by the study, the authors managed to ultimately run
only 102 (less than 20\%).
These low numbers only reflect the authors' success at running the code.
They did not check for correctness of the code (i.e., does the code actually implement what is advertised in the paper), nor the reproducibility of the results (does each run lead to the same results as in the paper).
One example of this problem can be found in \citet{topalidou:2015a}, in which the authors tried to replicate results obtained from a computational neuroscience model.
Source code was not available, neither as supplementary material to the paper nor in a public repository.
When the replicators obtained the source code after contacting the corresponding author, they found that it could not be compiled and would be difficult to reuse for other purposes.\\

Confronted with this problem, a small but growing number of journals and
publishers have reacted by adopting explicit policies for data and software. Examples can be seen in the PLOS instructions on
\href{http://journals.plos.org/plosone/s/materials-and-software-sharing}{Materials
  and Software Sharing} and
on \href{http://journals.plos.org/plosone/s/data-availability}{Data Availability},
and in the
\href{https://elifesciences.org/elife-news/inside-elife-forking-software-used-elife-papers-github}{recent
  announcement} by eLife on forking (creating a linked copy of) software used in eLife papers to GitHub.
Such policies help to ensure access to code and data in a well-defined format
\citep{perkel:2016} but this will not guarantee reproducibility nor
correctness.
At the educational and methodological levels, things have started to change with a
growing literature on best practices for making computations reproducible
\citep{sandve:2013, crook:2013, wilson:2014, halchenko:2015, janz:2015,
  hinsen:2015}.
  Related initiatives such as Software and Data Carpentry
\citep{wilson:2016} are of note since their goal is {\em to make
scientists more productive, and their work more reliable, by teaching them basic computing skills}.
Such best practices could be applied to already published research codebases as well, provided the original authors are willing to take on the challenge of 
re-implementing their software for the sake of better science.
Unfortunately, this is unlikely since the
incentives for doing such time-consuming work are low or nonexistent.
Furthermore, if the original authors made mistakes in their original
implementation, it seems likely that they will reproduce their mistakes
in any re-implementation.\\


\section*{Replication and reproduction}
\label{sec:replication-reproduction}
While recognition of the replication crisis as a problem
for scientific research has increased over time, unfortunately no 
common terminology has emerged so far. One reason for
the diverse use of terms is that each field of research has its own
specific technical and social obstacles on the road to publishing
results and findings that can be verified by other scientists. Here we
briefly summarize the obstacles that arise from the use of computers
and software in scientific research, and introduce the terminology we
will use in the rest of this article. We note, however, that there is
some disagreement about this particular choice of terminology even
among the authors of this article. \\

\textit{Reproducing} the result of a computation means running the
same software on the same input data and obtaining the same results.
The goal of a reproduction attempt is to verify that the computational
protocol leading to the results has been recorded correctly.
Performing computations reproducibly can be seen as a form of
provenance tracking, the software being a detailed record of all data
processing steps.

In theory, computation is a deterministic process and exact
reproduction should therefore be trivial. In reality, it is very
difficult to achieve because of the complexity of today's software
stacks and the tediousness of recording all interactions between a
scientist and a computer 
\citep[although a number of recent tools have attempted to automate such recording, e.g.][]{guo:2011, davison:2012, murta:2015}.
\citeauthor{Mesnard:2016} explain
\citep{Mesnard:2016} how difficult it can be to reproduce a two-year-old
computation even though all possible precautions were taken at the
time to ensure reproducibility.  The most frequent obstacles are the
loss of parts of the software or input data, lack of a computing
environment that is sufficiently similar to the one used initially,
and insufficient instructions for making the software work. An
obstacle specific to numerical computations is the use of
floating-point arithmetic, whose rules are subject to slightly
different interpretations by different compilers and runtime support
systems. A large variety of research practices and support tools have
been developed recently to facilitate reproducible computations. For a
collection of recipes that have proven useful, see
\citet{kitzes:2017}.

Publishing a reproducible computational result implies publishing all
the software and all the input data, or references to previously
published software and data, along with the traditional article
describing the work. An obvious added value is the availability of the
software and data, which helps readers to gain a better understanding
of the work, and can be re-used in other research projects. In
addition, reproducibly published results are more trustworthy, because
many common mistakes in working with computers can be excluded:
mistyping parameter values or input file names, updating the software
but forgetting to mention the changes in the description of the
method, planning to use one version of some software but actually
using a different one, etc.

Strictly speaking, reproducibility is defined in the context of
identical computational environments. However, useful scientific
software is expected to be robust with respect to certain changes in
this environment. A computer program that produces different results
when compiled using different compilers, or run on two different
computers, would be considered suspect by most practitioners, even if
it were demonstrably correct in one specific environment. Ultimately
it is not the software that is of interest for science, but the models
and methods that it implements. The software is merely a vehicle to
perform computations based on these models and methods. If results
depend on hard-to-control implementation details of the software,
their relation to the underlying models and methods becomes unclear
and unreliable.\\

\textit{Replicating} a published result means writing and then running
new software based on the description of a computational model or
method provided in the original publication, and obtaining results
that are similar enough to be considered equivalent. What exactly
``similar enough'' means strongly depends on the kind of computation
being performed, and can only be judged by an expert in the field.
The main obstacle to replicability is an incomplete or imprecise
description of the models and methods.

Replicability is a much stronger quality indicator than
reproducibility. In fact, reproducibility merely guarantees that all
the ingredients of a computation are well documented. It does not
imply that any of them are correct and/or appropriate for implementing
the models and methods that were meant to be applied, nor that the
descriptions of these models and methods are correct and clear. A
successful replication shows that two teams have produced independent
implementations that generate equivalent results, which makes serious
mistakes in either implementation unlikely. Moreover, it shows that the
second team was able to understand the description provided by the
first team.

Replication can be attempted for both reproducible and
non-reproducible results. However, when an attempt to replicate
non-reproducible work fails, yielding results too different to be
considered equivalent, it can be very difficult to identify the cause
of the disagreement. Reproducibility guarantees the existence of a
precise and complete description of the models and methods being
applied in the original work, in the form of software source code,
which can be analyzed during the investigation of any
discrepancies. The holy grail of computational science is therefore a
reproducible replication of reproducible original work.

\section*{The ReScience initiative}

Performing a replication is a daunting task that is traditionally not well
rewarded. Nevertheless, some people are willing to replicate computational
research. The motivations for doing so are very diverse (see Box 1). Students
may want to familiarize themselves with a specific scientific domain, and
acquire relevant practical experience by replicating important published
work. Senior researchers may critically need a specific piece of code for a
research project and therefore re-implement a published computational
method. If these people write a brand new open source implementation of already
published research, it is likely that this new implementation will be of
interest for other people as well, including the original authors. The question
is where to publish such a replication. To the best of our knowledge, no major
journal accepts replications in computational science for publication. This was
the main motivation for the creation of the ReScience journal
(\href{https://rescience.github.io}{rescience.github.io}) by Konrad Hinsen and
Nicolas P. Rougier in September 2015.\\

\begin{tcolorbox}[breakable, pad at break*=1mm,
                  colback=black!2.5, arc=0pt, outer arc=0pt, boxrule=.25pt]
\begin{footnotesize}
\textbf{Box 1.} Authors having  published in Rescience explain their motivation.\\

\textbf{(\cite{stachelek:2016})} I was motivated to replicate the results of
the original paper because I feel that working through code supplements to blog
posts has really helped me learn the process of scientific analysis. I could 
have published my replication as a blog post but I wanted the exposure and 
permanency that goes along with journal articles. This was my first experience 
with formal replication. I think the review was useful because it forced me to 
consider how the replication would be used by people other than myself. I have 
not yet experienced any new interactions following publication. However, I did 
notify the author of the original implementation about the replication's 
publication. I think this may lead to future correspondence. The original author suggested that he would consider submitting his own replications to
ReScience in the future.\\

\textbf{(\cite{topalidou:2015b})} Our initial motivation and the main reason
for replicating the model is that we needed it in order to collaborate with our
neurobiologist colleagues. When we arrived in our new lab, the model had just
been published (2013) but the original author had left the lab a few months before
our arrival. There was no public repository nor version control, and the paper
describing the model was incomplete and partly inaccurate. We managed to get
our hands on the original sources (6,000 lines of Delphi) only to realize we
could not compile them. It took us three months to replicate it using 250 lines
of Python. But at this time, there was no place to publish this kind of
replication to share the new code with colleagues. Since then, we have refined
the model and made new predictions that have been confirmed. Our initial
replication effort really gave the model a second life.\\

\textbf{(\cite{viejo:2016})} Replicating previous work is a relatively routine
task every time we want to build a new model: either because we want to build
on this previous work, or because we want to compare our new model to it. We also give replication tasks to
M.Sc. students every year, as projects. In all these cases, we are confronted
with incomplete or inaccurate model descriptions, as well as with the impossibility
to obtain the original results. Contacting the original authors sometimes
solves the problem, but not so often (because of the {\em dog ate my hard
  drive} syndrome). We thus accumulate knowledge, internal to the lab, about
which model works and which doesn't, and how a given model has to be parameterized
to really work. Without any place to publish it, this knowledge is
wasted. Publishing it in ReScience, opening the discussion publicly, will be a
progress for all of us. \par
\end{footnotesize}
\end{tcolorbox}

ReScience is an openly-peer-reviewed journal that targets computational research
and encourages the explicit replication of already published research. In order
to provide the largest possible benefit to the scientific community, replications are
required to be reproducible and open-source. In two years of existence, 17
articles have been published and 4 are currently under review
(\href{https://github.com/ReScience/ReScience-submission/pull/20}{\#20},
\href{https://github.com/ReScience/ReScience-submission/pull/27}{\#39},
\href{https://github.com/ReScience/ReScience-submission/pull/30}{\#41},
\href{https://github.com/ReScience/ReScience-submission/pull/30}{\#43}). The
editorial board covers a wide range of computational sciences (see
\url{http://rescience.github.io/board/}) and more than 70 volunteers have registered to be reviewers. The scientific domains of published work
are computational neuroscience, neuroimaging, computational ecology and
computer graphics, with a majority in computational neuroscience. The most
popular programming languages are Python and R. The review process takes about 100~days on average and involves about 50~comments. There is a
strong bias towards successful replication (100\%); experience has
taught us that researchers are reluctant to publish failed replications, even when they can prove that the original work is wrong. For young researchers,
there is a social/professional risk in publishing articles that show
results from a senior researcher to be wrong. Until we implement a
certified anonymized submission process, this strong bias will most likely
remain.\\

One of the specificities of the ReScience journal is a publishing
chain that is radically different from any other traditional
scientific journal, since ReScience lives on GitHub, a platform
originally designed for collaborative software development. A
ReScience submission is treated very similarly to a contribution to an
Open Source software project. One of the consequences is that the
whole process, from submission via reviewing
to publication, is open for anyone to see and even comment on.\\

Each submission is considered by a member of the editorial board, who
may decide to reject the submission if it does not respect the formal
publication criteria of ReScience. A submission must contain
\begin{itemize}
\item a precise reference to the work being replicated,
\item an explanation of why the authors think they have replicated the paper
  (same figures, same graphics, same behavior, etc.) or why they have failed.
\item a description of any difficulties encountered during the
      replication,
\item open-source code that produces the replication results,
\item an explanation of this code for human readers.
\end{itemize}
A complete submission therefore consists of both computer code and an
accompanying article, which are sent to ReScience in the form of a pull
request (the process used on GitHub to submit a proposed modification
to a software project).
Partial replications that cover only some of the results in the
original work are acceptable, but must be justified.\\

If the submission respects these criteria, the editor assigns it to
two reviewers for further evaluation and tests. The reviewers evaluate
the code and the accompanying material in continuous interaction with
the authors through the discussion section until both reviewers
consider the work acceptable for publication. The goal of the review
is thus to help the authors meet the ReScience quality standards
through discussion. Since ReScience targets replication of already
published work, the criteria of importance or novelty applied by most
traditional journals are irrelevant.

For a successful submission (i.e. partial or full replication) to be accepted,
both reviewers must consider it reproducible and a valid replication of the
original work. As we explained earlier, this means that the reviewers
\begin{itemize}
\item are able to run the proposed implementation on their computers,
\item obtain the same results as indicated in the accompanying paper,
\item consider these results sufficiently close to the ones reported in the
  original paper being replicated.
\end{itemize}

For a failure to replicate submission to be accepted, we require extra steps
to be taken. In addition to scrutiny of the submission by reviewers and
editors, we will try to contact the authors of the original research, and issue
a challenge to the community to spot and report errors in the new implementation.
If no errors are found, the submission will be accepted and the original
research will be declared non-replicable.

%

Since independent implementation is a major feature of replication
work, ReScience does not allow authors to submit replications of their
own research, nor the research of close collaborators. Moreover,
replication work should be based exclusively on the originally
published paper, although exceptions are admitted if properly documented
in the replication article. Mistakes in the
implementation of computational models and methods are often due to
biases that authors invariably have, consciously or not. Such biases
will inevitably carry over to a replication. Perhaps even more
importantly, cross-fertilization is generally useful in research, and
trying to replicate the work of one’s peers might pave the way for a
future collaboration, or may give rise to new ideas as a result of the
replication effort.

\section*{Lessons learned}

Although ReScience is still a young project, the submissions handled
so far already provide valuable experience concerning the
reproducibility and replicability of computational work in scientific
research.

\subsection*{Short-term and long-term reproducibility}

While some of the reasons for non-reproducibility are specific to each
scientific domain, our experience has shown that there are also some
common issues that can be identified.  Missing code and/or data,
undocumented dependencies, and inaccurate or imprecise description
appear to be characteristic of much non-reproducible work. Moreover,
these problems are not always easy to detect even for attentive
reviewers, as we discovered when some articles published in ReScience
turned out to be difficult to reproduce for someone else for exactly
the reasons listed above. ReScience reviewers are scientists working
in the same domain as the submitting authors, because familiarity with
the field is a condition for judging if a replication is
successful. But this also means that our reviewers share a significant
common background with the authors, and that background often includes
the software packages and programming languages adopted by their
community.  In particular, if both authors and reviewers have
essential libraries of their community installed on their computers,
they may not notice that these libraries are actually dependencies of
the submitted code.  While solutions to this problem evidently exist
(ReScience could, for example, request that authors make their
software work on a standard computational environment supplied in the
form of a virtual machine), they represent an additional effort to
authors and therefore discourage them from submitting replication work
to ReScience. Moreover, the evaluation of \textit{de-facto}
reproducibility (``works on my machine'') by reviewers is useful as
well, because it tests the robustness of the code under small
variations in the computational environments that are inevitable in
real life. Our goal is to develop a set of recommendations for authors
that represent a workable compromise between reproducibility,
robustness, and implementation effort. These recommendations will
evolve over time, and we hope that with improving technology we will
ultimately reach full reproducibility over a few decades.\\

Another issue with reproducibility is that with today's computing
technology, long-term reproducibility can only be achieved by imposing
drastic constraints on languages and libraries that are not compatible
with the requirements of research computing. This problem is nicely
illustrated by \citet{Mesnard:2016} whose authors report trying to
reproduce their own work performed two years earlier. Even though
Barba's group is committed to reproducible research practices, they
did not escape the many problems one can face when trying to re-run a
piece of code. As a consequence, code that is written for ReScience
today will likely cease to be functional at some point in the future.
The long-term value of a ReScience publication lies not just in the
actual code but also in the accompanying article. The combination of the
original article and the replication article provide a complete and
consistent description of the original work, as evidenced by the fact
that replication was possible. Even 5, 10, or 20 years later, a
competent scientist should be able to replicate the work again thanks
to these two articles. Of course, the new code can also help, but the
true long-term value of a replication is the accompanying article.\\

\subsection*{Open reviewing}

The well-known weaknesses of the traditional anonymous peer-reviewing
system used by most scientific journals have motivated many
experiments with alternative reviewing processes. The variant
adopted by ReScience is similar to the ones used by F1000Research or
PeerJ, but is even more radically open: anyone can look at ReScience
submissions and at the complete reviewing process, starting from the
assignment of an editor and the invitation of reviewers. Moreover,
anyone with a GitHub account can intervene by commenting. Such
interventions could even be anonymous because a GitHub account is not
required to advertise a real name or any other identifying
element. ReScience does currently require all authors, editors, and
reviewers to provide real names (which however are not verified in any
way), but there are valid reasons to allow anonymity for authors and
reviewers, in particular to allow junior scientists to criticize the
work of senior colleagues without fear of retribution, and we envisage
exploring such options in the future.

Our experience with this open reviewing system is very positive so
far. The exchanges between reviewers and authors are constructive and
courteous, without exception. They are more similar in style to a
coffee-table discussion than to the judgement/defence style that
dominates traditional anonymous reviewing. Once reviewers have been
invited and have accepted the task, the editors' main role is to
ensure that the review moves forward, by gently reminding everyone to
reply within reasonable delays. In addition, the editors occasionally
answer questions by authors and reviewers about the ReScience
publishing process.

The possibility to involve participants beyond the traditional group of
authors, editors, and reviewers is particularly interesting in the
case of ReScience, because it can be helpful to solicit input from the
authors of the original study that is being replicated. For example,
in one recent case
(\href{https://github.com/ReScience/ReScience-submission/pull/28}{\#28}),
a reviewer suggested asking the author of the original work for
permission to re-use an image. The author intervened in the review and
granted permission.

\subsection*{Publishing on the GitHub platform}

\href{http://github.com/}{GitHub} is a commercial platform for collaborative
software development based on the popular version control system
\href{https://git-scm.com/}{git}. It offers unlimited free use to public
projects, defined as projects whose contents are accessible to everyone. All
ReScience activities are organized around a few such Open Source projects
hosted by GitHub. This is an unusual choice for a scientific journal, the only
other journal hosted on GitHub being The Journal of Open Source
Software \citep{Smith:2017}. In this section, we discuss the advantages and
problems resulting from this choice, considering both technical and social
issues.

There are clear differences between platforms for software
development, such as GitHub, and platforms for scientific publishing,
such as \href{http://home.highwire.org/}{HighWire}. The latter tend to
be expensive commercial products developed for the needs of large
commercial publishers, although the market is beginning to diversify
with products such as
\href{https://www.episciences.org/}{Episciences}.  More importantly,
to the best of our knowledge, no existing scientific publishing
platform supports the submission and review of code, which is an
essential part of every ReScience article. For this reason, the only
option for ReScience was to adopt a software development platform and
develop a set of procedures that make it usable for scientific
publishing.

Our experience shows that the GitHub platform provides excellent
support for the reviewing process, which is not surprising given that the
review of a scientific article containing code is not fundamentally
different from the review of code with accompanying documentation.
One potential issue for other journals envisaging adoption of this
platform is the necessity that submitting authors have a basic
knowledge of the version control system Git and of the techniques of
collaborative software development. Given the code-centric nature of
ReScience, this has not been a major problem for us, and the minor
issues have been resolved by our editors providing technical assistance
to authors. It is of course possible that potential authors are
completely discouraged from submitting to ReScience by their lack
of the required technical competence, but so far nobody has
provided feedback suggesting that this is a problem.

The main inconvenience of the GitHub platform is its almost complete
lack of support for the publishing steps, once a submission has
successfully passed the reviewing process. At this point, the
submission consists of an article text in Markdown format plus a set
of code and data files in a git repository. The desired archival form
is an article in PDF format plus a permanent archive of the submitted
code and data, with a Digital Object Identifier (DOI) providing a
permanent reference. The \href{https://zenodo.org/}{Zenodo} platform
allows straightforward archiving of snapshots of a repository hosted
on GitHub, and issues a DOI for the archive. This leaves the task
of producing a PDF version of the article, which is currently handled
by the managing editor of the submission, in order to ease the technical
burden on our authors.

A minor inconvenience of the GitHub platform is its implementation of
code reviews. It is designed for reviewing contributions to a
collaborative project. The contributor submits new code and
modifications to existing code in the form of a ``pull request'',
which other project members can then comment on. In the course of the
exchanges, the contributor can update the code and request further
comments. Once everybody is satisfied, the contribution is ``merged''
into the main project. In the case of ReScience, the collaborative
project is the whole journal, and each article submission is a
contribution proposed as a pull request. This is, however, not a very
intuitive representation of how a journal works. It would be more
natural to have a separate repository for each article, an arrangement
that would also facilitate the final publishing steps. However, GitHub
does not allow code review on a new repository, only on contributions
to an already existing one.

Relying on a free-use offer on a commercial platform poses some
additional problems for scientific publishing. GitHub can change its
conditions at any time, and could in principle delete or modify
ReScience contents at any time without prior notice. Moreover, in the
case of technical problems rendering ReScience contents temporarily or
permanently inaccessible, the ReScience community has no legal claims
for compensation because there is no contract that would imply any
obligations for GitHub. It would clearly be imprudent to count on
GitHub for long-term preservation of ReScience content, which is why
we deposit accepted articles on Zenodo, a platform designed for
archiving scientific information and funded by research organizations
as an element of public research infrastructure.

The use of free services provided by GitHub and Zenodo was clearly
important to get ReScience started. The incentives for the publication
of replication work being low, and its importance being recognized
only slowly in the scientific community, funding ReScience through
either author page charges or grants would have created further
obstacles to its success. A less obvious advantage of not having to
organize funding is that ReScience can exist without being backed by
any legal entity that would manage its budget. This makes it possible
to maintain a community spirit focused on shared scientific
objectives, with nobody in a position to influence ReScience by
explicit or implicit threats of reducing future funding.

\begin{figure}
  \includegraphics[width=1.0\columnwidth]{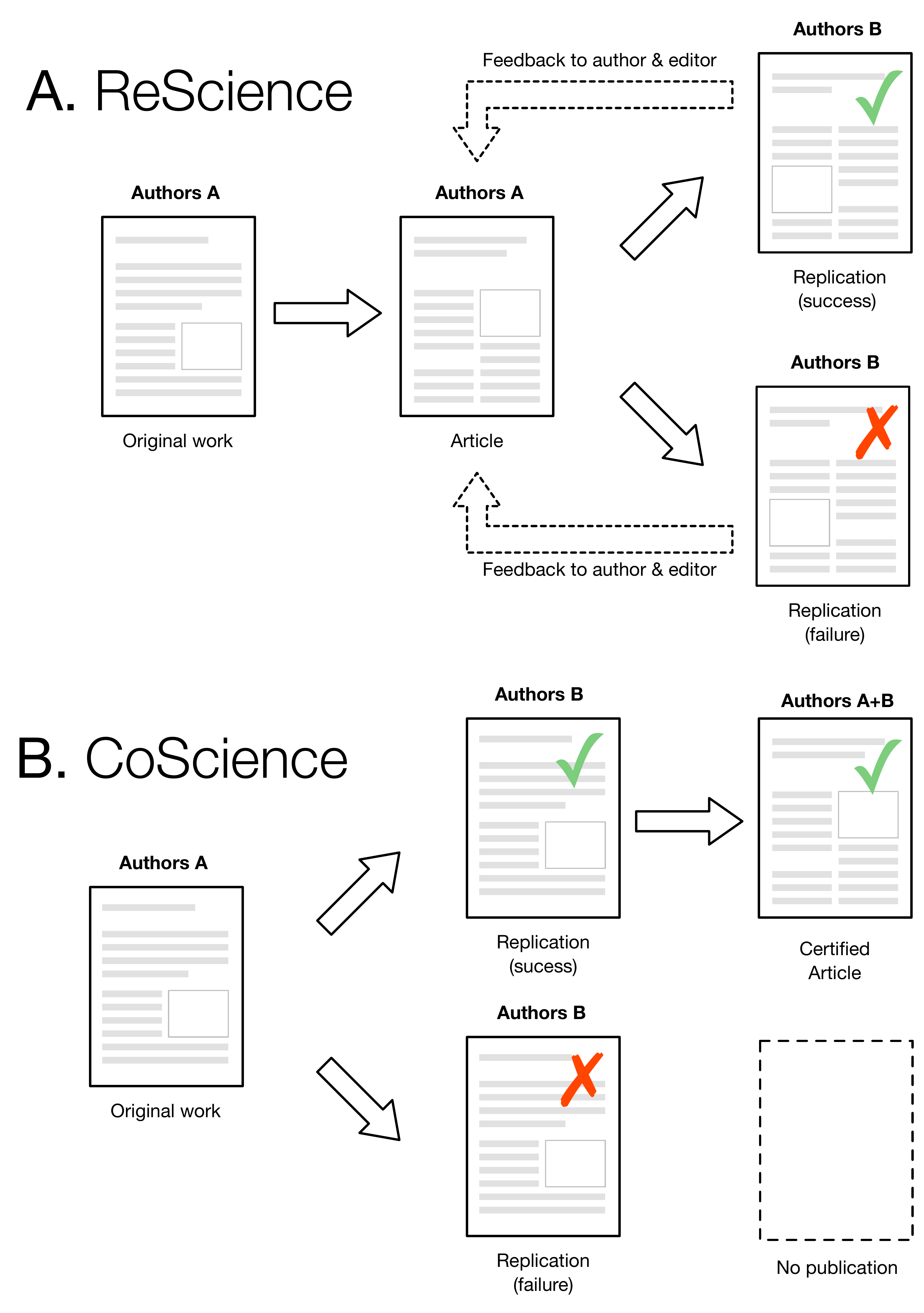}
  \caption{\textbf{A} The ReScience publication chain starts from an
    original research article by authors A, published in a journal, in
    conference proceedings, or as a preprint. This article constitutes
    the base material for authors B, who attempt to replicate the work
    based on its description. Success or failure to replicate is not a
    criterion for acceptance or rejection, even though failure to
    replicate requires more precaution to ensure this is not a
    misunderstanding or a bug in the new code. After review, the
    replication is published, and feedback is given to original
    authors (and editors) to inform them the work has been replicated
    (or not). \textbf{B} The CoScience proposal would require the
    replication to happen \textit{before} the actual publication. In
    case of failure, nothing will be published. In case of success,
    the publication will be endorsed by authors A and authors B with
    identified roles and will be certified as reproducible because it
    has been replicated by an independent group.}
  \label{fig:coscience}
\end{figure}

\section*{Outlook}

Based on our experience with the ReScience initiative, we can engage
in informed speculation about possible future evolutions in scientific
publishing, in particular concerning replication work. We will not
discuss minor technical advances such as a better toolchain for
producing PDF articles, but concentrate on long-term improvements in
the technology of electronic publishing and, most of all, in the
attitude of the scientific community towards the publication,
preservation, and verification of computer-aided research.


A fundamental technical issue is the difficulty of archiving or
accurately describing the software environments in which computational
scientists perform their work. A publication should be accompanied by
both a human-readable description of this environment and an
executable binary form. The human-readable description allows an
inspection of the versions of all software packages that were used,
for example to check for the impact of bugs that become known only
after a study was published. The executable version enables other
scientists to re-run the analyses and inspect intermediate
results. Ideally, the human-readable description would permit
rebuilding the executable version, in the same way that software
source code permits rebuilding executable binaries. This approach is
pursued for example by the package manager Guix \citep{Courtes:2015}.
A more limited but still useful implementation of the same idea exists
in the form of the conda package manager \citep{conda}, which uses a
so-called environment file to describe and reconstruct environments.
The main limitation compared to Guix is that the packages that make up
a conda environment are themselves not reproducible. For example, a
conda environment file does not state which compiler versions were
used to build a package.

Containerization, as implemented e.g. by Docker \citep{docker} is
currently much discussed, but provides only the executable version
without a human-readable description. Moreover, the long-term
stability of the container file format remains to be
evaluated. History has shown that long-term stability in computing
technology is achieved only by technology for which it is a design
priority, as in the case of the Java Virtual Machine
\citep{JVM}. Docker, on the contrary, is promoted as a deployment
technology with no visible ambition towards archiving of computational
environments.


Today's electronic publishing platforms for scientific research still
show their origins in paper-based publishing. Except for the
replacement of printed paper by a printable PDF file, not much has
changed. Although it is increasingly realized that software and data
should be integral parts of most scientific publications today, they
are at best relegated to the status of ``supplementary material'', and
systematically excluded from the peer review process. In fact, to the
best of our knowledge, ReScience is the only scientific journal that
aims to verify the correctness of scientific software. As our
experience has shown, it is far easier to graft publication onto a
software development platform than to integrate software reviewing
into a publishing platform. Furthermore, tools that will allow for the automated
validation of computational models and the automated verification of correctness are being actively 
developed in the community (see, for example, 
\href{https://github.com/scidash/sciunit}{SciUnit} or 
\href{https://github.com/OpenSourceBrain/osb-model-validation}{OSB-model-validation}). 
An integration of such frameworks, which would greatly enhance the 
verification and validation process, seems feasible for the existing software 
development platforms.

A logical next step is to fully embrace the technology designed for
software development, which far better takes into account the
specificity of electronic information processing than today's
scientific publishing systems. In addition to the proper handling of
code, such an approach offers further advantages. Perhaps the most
important one is a shift of focus from the paper as a mostly isolated
and finished piece of work to scientific progress as a collection of
incremental and highly interdependent steps. The
\href{https://www.softwareheritage.org/}{Software Heritage} project,
whose aim is to create a permanent public archive of all publicly
available software source code, adopts exactly this point of view for
the preservation of software. As our experience with ReScience has
shown, integrating the narrative of a scientific article into a
framework designed for software development is not difficult at all.
Publishing and archiving scientific research in Software Heritage
would offer several advantages. The intrinsic identifiers that provide
access to the contents of the archive permit unambiguous and permanent
references to ongoing projects as well as to snapshots at a specific
time, and to whole projects as well as to the individual files that
are part of them. Such references hold the promise for better reuse of
scientific information, for better reproducibility of
computations, and for fairer attribution of credit to scientists
who contribute to research infrastructure.


One immediate and legitimate question is to wonder to what extent a
replication could be performed \textit{prior} to the publication of
the original article. This would strongly reinforce a claim because a
successful and independent replication would be available right from
the start. As illustrated in Figure~\ref{fig:coscience}, this would
require group A to contact group B and send them a draft of
their original work (the one that would be normally submitted to a
journal) such that group B could perform a replication and confirm or
refute the results. In case of confirmation, a certified article could
be later published with both groups as authors (each group being
identified according to their respective roles). However, if the
replication fails and the original work cannot be fixed, this would
prevent publication. This model would improve the quality of
computational research and also considerably slow down the rapid pace
of publication we are observing today. Unfortunately, such a scenario
seems highly improbable today. The pressure to publish is so strong
and the incentive for doing replication so low that it would most
probably prevent such collaborative work. However, we hope that the
current replication crisis will lead to a change in attitude, with
an emphasis on the quality rather than the quantity of scientific ouput,
with CoScience becoming the gold-standard approach to quality assurance.


\renewcommand*{\bibfont}{\footnotesize}
\printbibliography[title=References]

\end{document}